\documentclass[12pt]{article}
\usepackage{amsmath}
\usepackage{amssymb}
\usepackage{euscript}

\usepackage{epsfig}

\usepackage{cite}

\usepackage{alltt}

\usepackage{epic}

\def\lint{\int\limits}
\def\bbeta{\bar{\beta}}

\newcommand{\Rcal}{{\cal R}}

\begin{document}                     

\allowdisplaybreaks

\begin{titlepage}
\begin{flushright}
{\bf  CERN-TH/2002-195\\
      UTHEP-02-0802 \\
    MPI-PhT-2002-036 }
\end{flushright}

\begin{center}
{\Large\bf
Electric Charge Screening Effect in\\
Single-W Production with the KoralW Monte Carlo$^\star$ 
}
\end{center}

\begin{center}
  {\bf   S. Jadach$^{a}$,}
  {\bf   W. P\l{}aczek$^{b,c}$,}
  {\bf   M. Skrzypek$^{a,c}$,}
  {\bf   B.F.L. Ward$^{d,e,c}$}
  {\em and}
  {\bf   Z. W\c{a}s$^{a,c}$ } \\
{\em $^a$Institute of Nuclear Physics,
  ul. Radzikowskiego 152, 31-342 Cracow, Poland}\\
{\em $^b$  Institute of Computer Science, Jagellonian University,\\
        ul. Nawojki 11, Cracow, Poland}\\
{\em $^c$CERN, Theory Division, CH-1211 Geneva 23, Switzerland}\\
{\em $^d$Department of Physics and Astronomy,\\
  The University of Tennessee, Knoxville, TN 37996-1200, USA}\\
{\em $^e$Werner-Heisenberg-Institut, Max-Planck-Institut f\"ur Physik,
        F\"ohringer Ring 6,\\ 80805 Munich, Germany }
\end{center}

\begin{abstract}
Any Monte Carlo event generator in which only initial state radiation (ISR)
is implemented, or ISR is simulated independently of the final state
radiation (FSR),
may feature too many photons with large transverse momenta, which deform
the topology of events
and result in too strong an overall energy loss due to ISR.
This {\em overproduction} of ISR photons happens in the presence
of the final state particle close to the beam particle of the same 
electric charge.
It is often said that the lack of the electric charge screening  effect
between ISR and FSR is responsible for the above pathology in ISR.
We present an elegant approximate method of curing the above problem,
without actually reinstalling FSR.
The method provides theoretical predictions of  modest 
precision: $\leq 2\%$. It is,
however, sufficient for the current $1W$ data analysis at the LEP2 collider.
Contrary to alternative methods implemented in other MC programs,
our method works for the ISR multiphotons with finite $p_T$.
Although this method is not an exact implementation of the complete/exact
ISR, FSR and their interference, it is very closely modelled on it.
We present a variety of numerical results obtained with
the newest version of the {KoralW} Monte Carlo, in which
this method is already implemented.
\end{abstract}

\vspace{-1mm}
\begin{center}
{\it To be submitted to European Physical Journal C}
\end{center}

\footnoterule
\noindent
{\footnotesize
\begin{itemize}
\item[${\star}$]
Work supported in part by the Polish Government grants KBN 5P03B09320,
2P03B00122, 
the NATO Grant PST.CLG.977751,
the US DoE contract DE-FG05-91ER40627,
the European Commission 5-th framework contract HPRN-CT-2000-00149
and the Polish-French Collaboration within IN2P3 through LAPP Annecy.
\end{itemize}
}
\begin{flushleft}
{\bf CERN-TH/2002-195\\
     September 2002 
}
\end{flushleft}
\end{titlepage}

\section{Introduction}
In the Monte Carlo programs for $e^-e^+$ colliders, one has to
model initial state radiation (ISR) as precisely as possible.
For some processes where the precision requirements are moderate,
such as the single-$W$ production process at LEP,
where the combined LEP2 precision is estimated at $\sim 7\%$, 
see Ref.~\cite{ICHEP2002},
it is sufficient to implement
ISR in the leading-logarithmic (LL) approximation.
This can be done by modelling ISR in the collinear, $p_T=0$,
approximation (see for instance refs.~\cite{Pukhov:1999gg,Montagna:2000ev}),
by ``unfolding'' the strictly collinear structure functions 
\cite{Berends:2000fj}
or using the so-called ``parton shower'' technique as in 
Ref.~\cite{Kurihara:2001} (recently employed also in \cite{Accomando:2002sz}),
or the Yennie--Frautschi--Suura (YFS) exclusive exponentiation
employed in Ref.~\cite{koralw:2001};
see Ref.~\cite{4f-LEP2YR:2000} for a more complete list of references.
In all of the above MC programs the final state radiation (FSR) from 
charged final particles
is either not modelled at all or simulated completely independently of
the ISR,
which is the general spirit of the LL approximation.
This may be problematic if one of the final state charged particles gets
close to the beam particle with the same electric charge.
In such a case ISR photon emission is damped, because of the well-known
{\em electric charge screening} effect (ECS),
and the total energy distribution of the ISR photons gets softer.
In the terminology of the LL it is described as the ``decrease of 
the LL scale''
from $s$, which is the square of the centre-of-mass-system (CMS) energy,
to $t$, which is the four-momentum transfer (squared)
from the beam to the nearby particle of the same charge.
We shall refer to the above phenomenon as 
the {\em leading-logarithmic scale transmutation} (LLST).

Let us stress that in the calculations based from the beginning on the
YFS exclusive exponentiation, as in the BHWIDE~\cite{bhwide:1997}
or KKMC~\cite{Jadach:1999vf,Jadach:2000ir} MC programs,
the ECS and LLST are automatically built in to an infinite order,
hence there is no need to reinstall them.
The  ECS and LLST in exclusive exponentiation are direct manifestations of the 
interference effects between ISR and FSR. 
Hence, they are also often described as ``coherence effects''. 

As already said, owing to the lack of ECS in the Monte Carlo 
calculation in which ISR
is modelled using the $s$-scale, one may encounter two pathologies:
(a) overproduction of photons with large transverse momenta,
which results in the noticeable deformation of the topology of events and
(b) too strong an overall energy loss due to ISR.
In particular this happens in the single-$W$ ($1W$)
production process at LEP2, when modelled
by the MC program KoralW~\cite{koralw:1995a,koralw:1995b,koralw:1998}.
The aim of this work is to invent a simple method of correcting for ECS,
and curing the above deficiencies, without the need of generating FSR in the MC%
\footnote{An obviously better alternative is to implement YFS exclusive
 exponentiation for the entire $e^-e^+\to 4f$ process. This is however
 too much work to be completed before the end of the LEP2 data analysis.
 For the $e^-e^+\to WW\to 4f$ subprocess, where precision requirements
are $\sim 0.5\%$, 
 the ECS/coherence effects between ISR and FSR ($W$ decays) are strongly suppressed.}.
The method is quite general, although its most immediate application
is improving the theoretical prediction for the $1W$ production process at LEP2
to the $\leq 2\%$ precision level.
Contrary to alternative curing methods implemented in the other 
MC programs~\cite{Montagna:2000ev,Kurihara:2001},
our method works for finite-$p_T$ ISR multiphoton distributions.
It is not  an ``ad hoc'' method because its development is based on the consideration
of the exact YFS exclusive exponentiation
for ISR, FSR and their interference~\cite{Jadach:2000ir}.

After defining our method,
we shall also present a variety of numerical results exploiting
the newest version of the {KoralW} Monte Carlo, in which
this method of introducing ECS is already implemented.

The {\em electric charge screening} effect
together with {\em leading-logarithmic scale transmutation}
for the multiphoton emission are described in the following section. 
The proposed method is {\em practical} because:
\begin{itemize}
\item
  the effects are introduced by means of a simple well-behaving Monte Carlo weight,
\item
  it does not require generation of the additional photons in the final state (FSR),
\item
  it provides an ISR precision good enough for LEP2 data analysis, 
  that is $\leq 2\%$.
\end{itemize}
The construction of the ECS weight is first described in 
Section~\ref{susec:Mhamha},
using a simple example of the  process $\mu^-\mu^+ \to \mu^-\mu^+$, for which
the ECS effect and the method of its introduction in the ISR is illustrated
with the help of many examples of the photon distributions.
This method is later generalized to the $e^-e^+ \to 4f$ process
in Section~\ref{sec:ECSwt}, where also the relevant
correction to the virtual form factor (overall normalization) is defined.
Numerical cross-checks and final discussion are provided in Section\
\ref{sec:numerics}.

\subsection{Electric charge screening}
\label{susec:ECS}

The electric charge screening effect is a well-known phenomenon 
of suppression of the photon emission from any subset 
of compensating charges (charges adding to zero or to small total charge),
which are close in the momentum phase space,
i.e. which have small effective mass, small angular distance, etc.
One should exclude from the consideration  
charges/particles that are well {\em time-separated},
i.e. from the decay and production parts of the narrow resonance 
production/decay processes.
A good example of a subset of compensating charges is 
for instance a fast-moving light $\mu^+\mu^-$ pair 
-- a ``dipole'' of compensating charges.
For the compensation considerations,
electric charges of every initial state particle should be assigned
an additional minus sign with respect to  particle charges in the final state
(much as in the definition of the four-momentum transfer).
For instance in the low-angle Bhabha (LABH) process, the initial state $e^-_i$
and the final state $e^-_f$ form together a ``dipole''
of compensating charges and the ECS/LLST takes place, 
which means that the emission of photons with an angle greater than the 
angular size
of the $(e^-_i,e^-_f)$ compensating system gets strongly suppressed.

The leading-logarithmic scale transmutation is just another face of the same ECS.
For ISR taken alone, the LL parameter 
$L_e(s)= 2\frac{\alpha}{\pi} \ln\frac{s}{m_e^2}$, which
controls the strength of the loss of the total energy due to ISR,
is defined for the so-called LL energy scale $\sqrt{s}$,
being just the total centre-of-mass system (CMS) energy.
Actually this so-called big logarithm $L_e(s)$
comes from the integral over the photon angle
$L_e(s) \simeq \int^{\pi}_{m^2/s} \frac{d\vartheta}{\vartheta}$.
Taking into account the interference between ISR and FSR introduces ECS.
This becomes important when, for instance, one of the final state $e^-$'s approaches
the initial beam $e^-$ by an angle $\theta$.
Then, the photon angular distribution gets suppressed
(i.e. loses the $1/\vartheta$ dependence) for $\vartheta>\theta$.
Hence $L_e(s)$ gets replaced in the so-called LL ISR structure functions by
a smaller $L_e(t) \simeq \int^{|t|/s}_{m^2/s} \frac{d\vartheta}{\vartheta}$.
Obviously, LLST is a numerically important effect when $|t| \ll s$.
Such is the case of the $1W$ production, where at least one of the
final state electrons gets inside the beam pipe due to
the strong peak in its scattering angle
(because of $t$-channel photon exchange in the Born distribution).

All the above discussion was for real photons, and in fact the whole ECS
can be viewed as an almost completely {\em classical} phenomenon.
The real photon angular distribution, in the soft-limit approximation,
is just the square of the classical electric current vector.
Quantum-mechanical virtual corrections merely correct the overall
normalization, cancelling properly infrared (IR) singularities 
in the photon {\em energy} distribution.
(The IR-divergent part of the exponentiated virtual correction
has necessarily  $L_e(t)$ as a coefficient.)

\subsection{``Mhamha'' toy example}
\label{susec:Mhamha}

As already indicated, in our method of implementing the ECS effect in ISR
we do not include explicit FSR; however, we take as a guide the YFS 
model in which
ISR, FSR and their interference are included.
In this section we examine carefully the ECS effect for a single real photon
in such a complete model, proposing at the end of the exercise a simple correcting weight
introducing the ECS effect in the pure ISR case.

\begin{figure}[!ht]
\begin{center}
\epsfig{file=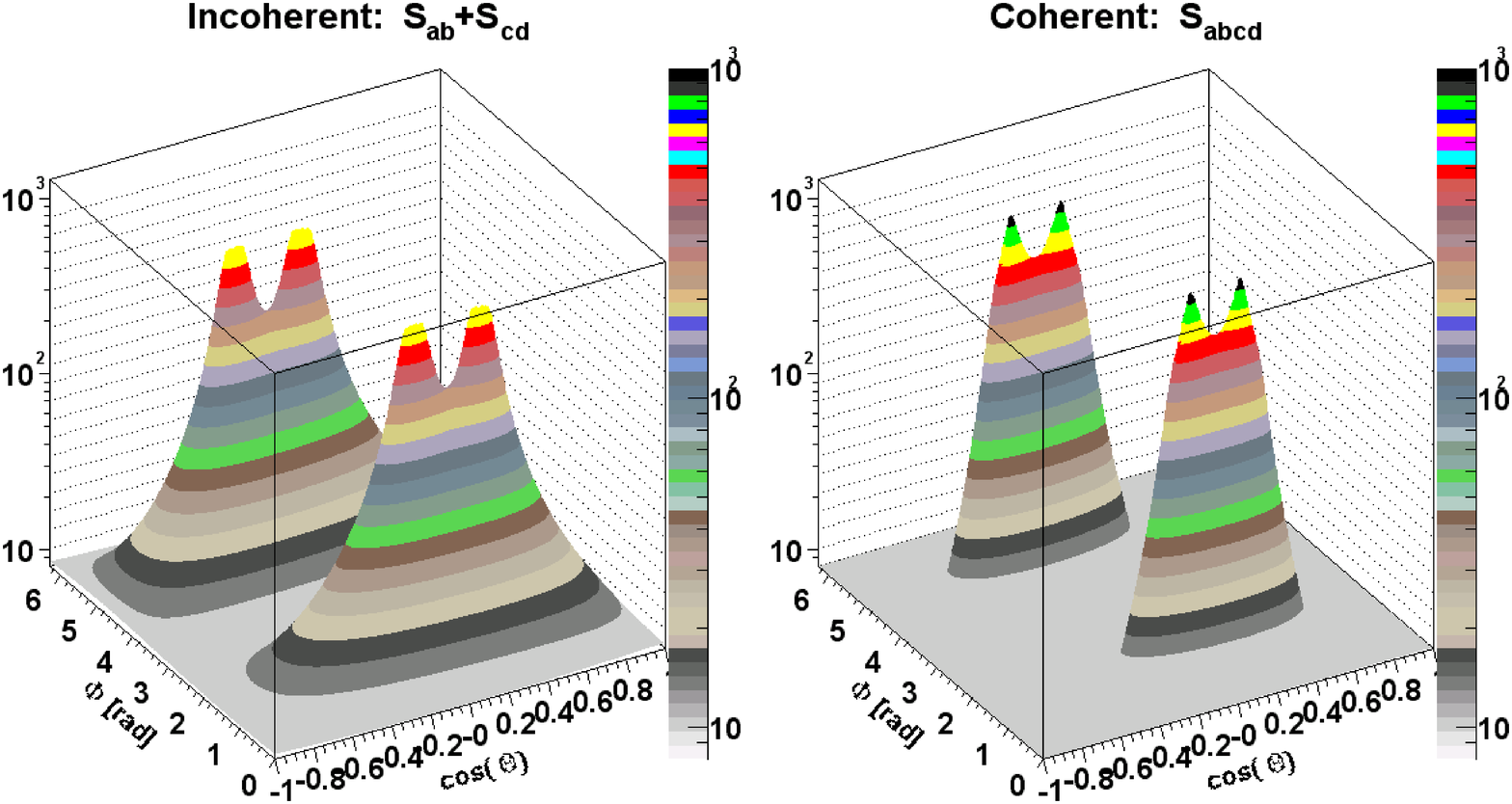,width=170mm,height=90mm}\\
\epsfig{file=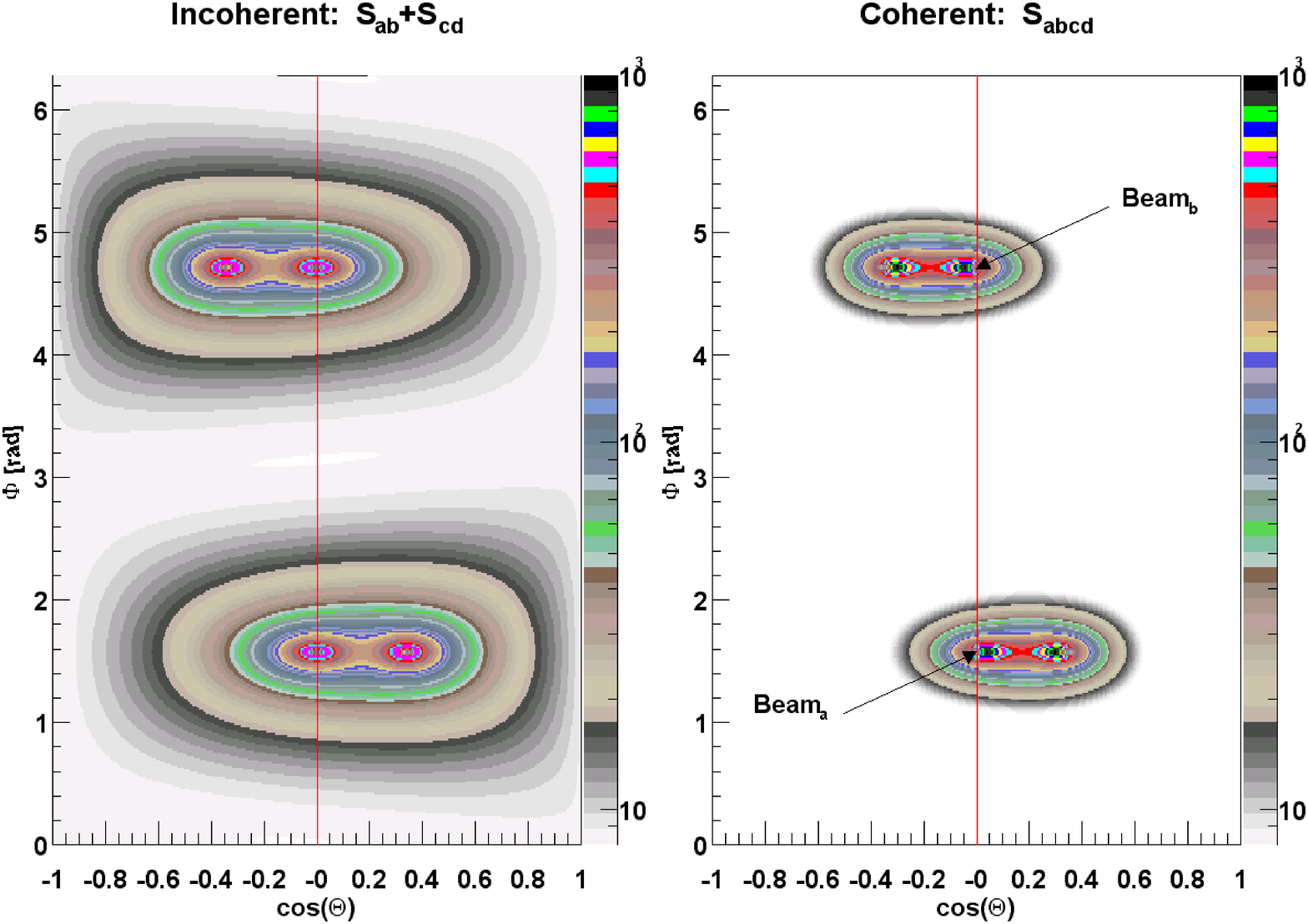,width=170mm,height=110mm}
\end{center}
\caption{\sf\small
  Photon angular distribution in ``Mhamha scattering'' 
$\mu^-\mu^+\to \mu^-\mu^+\gamma$ at $\sqrt{s}=5$ GeV and muon scattering
angle of 20$^\circ$. The difference between left-
  and right-hand side plots illustrates the electric charge screening effect.
}
\label{fig:Mhamha1}
\end{figure}

\begin{figure}[!ht]
\begin{center}
\epsfig{file=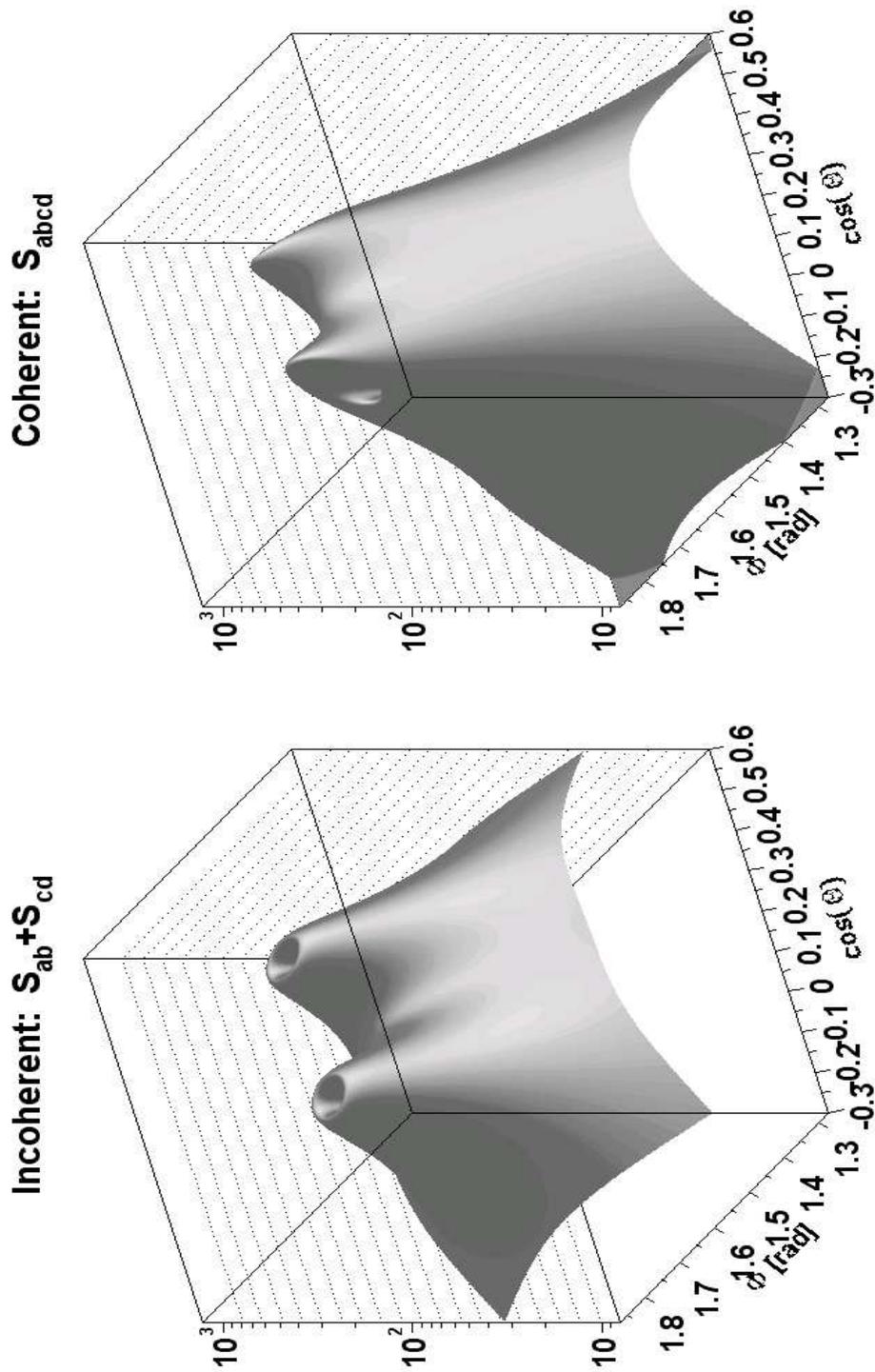,width=200mm,height=130mm,angle=90}
\end{center}
\caption{\sf\small
  Photon angular distribution. 
The same as in Fig.~\ref{fig:Mhamha1}.
  Range restricted to vicinity of a-c dipole. ``Craters'' are ``helicity
zeros''. 
}
\label{fig:Mhamha2}
\end{figure}

The above will be done by carefully scrutinizing several variants
of the angular photon distributions of a ``toy process'':
\begin{equation}
  \mu^-(p_a)+\mu^+(p_b)   \to  \mu^-(p_c)+\mu^+(p_d) +\gamma(k),
\end{equation}
a muonic analogue of the Bhabha process, which we shall call 
a ``Mhamha scattering process''%
\footnote{The name ``Mhamha'' was suggested by A. Blondel in the context of the
 muon collider studies.}.
In the soft photon limit, the final state distribution is equal to the 
Born distribution
times the {\em complete} soft photon factor
\begin{equation}
  \tilde{S}_{abcd} = -\frac{\alpha}{4\pi^2} 
  \left(  \frac{p_a}{k p_a} -\frac{p_b}{k p_b} -\frac{p_c}{k p_c} +\frac{p_d}{k p_d}   \right)^2,
\end{equation}
the same factor which is a basic element in the YFS exponentiation~\cite{yfs:1961}.
In order to see the ECS effect we compare the above complete $\tilde{S}_{abcd}$
with the incoherent sum $\tilde{S}_{ab}+\tilde{S}_{cd}$
of the ISR and FSR contributions, neglecting the interference IFI=ISR$\otimes$FSR,
where
\begin{equation}
  \tilde{S}_{ab} = -\frac{\alpha}{4\pi^2}
               \left(  \frac{p_a}{k p_a} -\frac{p_b}{k p_b} \right)^2,\quad
  \tilde{S}_{cd} = -\frac{\alpha}{4\pi^2}
               \left(  \frac{p_c}{k p_c} -\frac{p_d}{k p_d} \right)^2.
\end{equation}
We choose a relatively small scattering angle $\theta_{ab}$.
In Fig.~\ref{fig:Mhamha1} we plot 
$\tilde{S}_{ab}+\tilde{S}_{cd}$ and complete $\tilde{S}_{abcd}$
side by side, as a function of the photon polar variables $(\cos\vartheta,\varphi)$.
In order to see very clearly the structure of the distribution over the entire unit sphere,
we choose $(\cos\vartheta,\varphi)$ with respect to the $z$-axis
pointing 
perpendicularly
to the beam axis (beams are along the $x$-axis).
We choose $\sqrt{s}=5$ GeV at which muons are already very relativistic
and the muon scattering angle 20$^\circ$ -- small enough to see ECS effect.
The four peaks of the photon emission intensity in Fig.~\ref{fig:Mhamha1}
are centred around directions of the four charged muons.
The ECS effect is seen as a clear suppression of the photon emission intensity
beyond the two dipole peaks $(a,c)$ and $(b,d)$. 
The first dipole is for the incoming and outgoing $\mu^-$ and the second one
is for the incoming and outgoing $\mu^+$.
In Fig.~\ref{fig:Mhamha1} the two $t$-channel dipole peaks are much sharper for
the complete $\tilde{S}_{abcd}$ than for the incoherent superposition of 
ISR and FSR,
$\tilde{S}_{ab}+\tilde{S}_{cd}$.
The top part of the dipole peaks is worth inspection, see Fig.~\ref{fig:Mhamha2},
as it features the well-known ``helicity zeros'' as deep ``craters'',
exactly in the direction of the emitter charge.
The craters are of an angular size $m_\mu/E_{beam}$
  and will shrink to a negligible size at very high energies.
The ECS effect is also
very clearly seen again in Fig.~\ref{fig:Mhamha2},
 as a strong sharpening of the photon radiation
intensity beyond the dipole double peak.

\begin{figure}[!ht]
\begin{center}
\epsfig{file=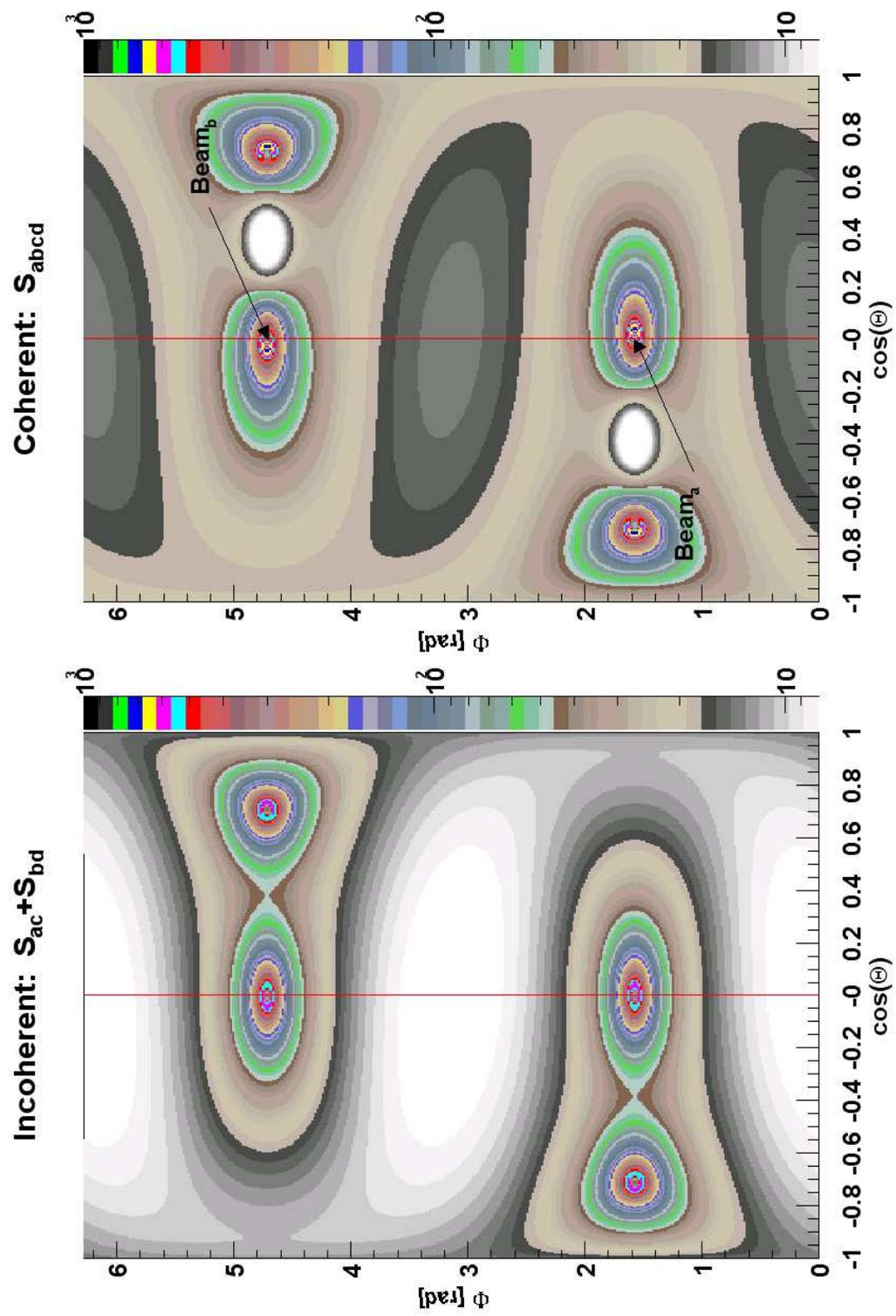,width=200mm,height=140mm,angle=90} 
\end{center}
\caption{\sf\small
  Photon radiation in backward scattering, $\theta=135^\circ$.
}
\label{fig:Mhamha3}
\end{figure}

Is the radiation pattern of $\tilde{S}_{abcd}$, which we see in Fig.~\ref{fig:Mhamha2},
specific to a $t$-channel character of the process?
Not really.
The radiation pattern of the real $\mu^-\mu^+$ pair with effective mass
equal to $\sqrt{|t|}$ 
and boosted to the total energy of 5 GeV is completely {\em undistinguishable}
from what we see in Fig.~\ref{fig:Mhamha2}.
The other way of describing it is that $\tilde{S}_{abcd}$ is extremely well approximated
by the incoherent sum of $\tilde{S}_{ac}+\tilde{S}_{bd}$.
Let us stress that $\tilde{S}_{ac}$ and $\tilde{S}_{bd}$ look very simple and natural
in the two Breit frames (rest frames of $(a,c)$ and $(b,d)$) and
the {\em non-trivial}
shape of the photon intensity in Fig.~\ref{fig:Mhamha2} is the result
of the {\em trivial} Lorentz boost from the Breit to the CMS frame.
(In a sense, the non-trivial shape plotted in Fig.~\ref{fig:Mhamha2}
is the ``fault'' of the observer himself, who has chosen to examine it
from an ``unnatural'' reference frame.)
The other interesting observation is that
if we have plotted $\tilde{S}_{ac}+\tilde{S}_{bd}$ in Figs.~\ref{fig:Mhamha1} and~\ref{fig:Mhamha2}
then we will have found that it is completely undistinguishable
from the $\tilde{S}_{abcd}$%
\footnote{This is why the BHWIDE MC program, 
which models $\tilde{S}_{ac}+\tilde{S}_{bd}$
  at the low MC level, is so efficient in modelling
  the multiphoton radiation in the complete wide-angle Bhabha process.}.
In order to see any noticeable difference between these two distributions,
we have to go to the backward scattering angle.
We do this in Fig.~\ref{fig:Mhamha3}, where we
compare these two emission distributions for the scattering angle $\theta=135^\circ$.
The peaking structure is the same; however, for $\tilde{S}_{abcd}$ we see
a rich interference pattern, especially for photons far away from the four charges.
At this large scattering angle 
we have checked that the $\tilde{S}_{ac}+\tilde{S}_{bd}$ and
$\tilde{S}_{ab}+\tilde{S}_{cd}$ distributions look almost identical.

\subsection{ECS weight -- preliminaries}
\begin{figure}[!ht]
\begin{center}
\epsfig{file=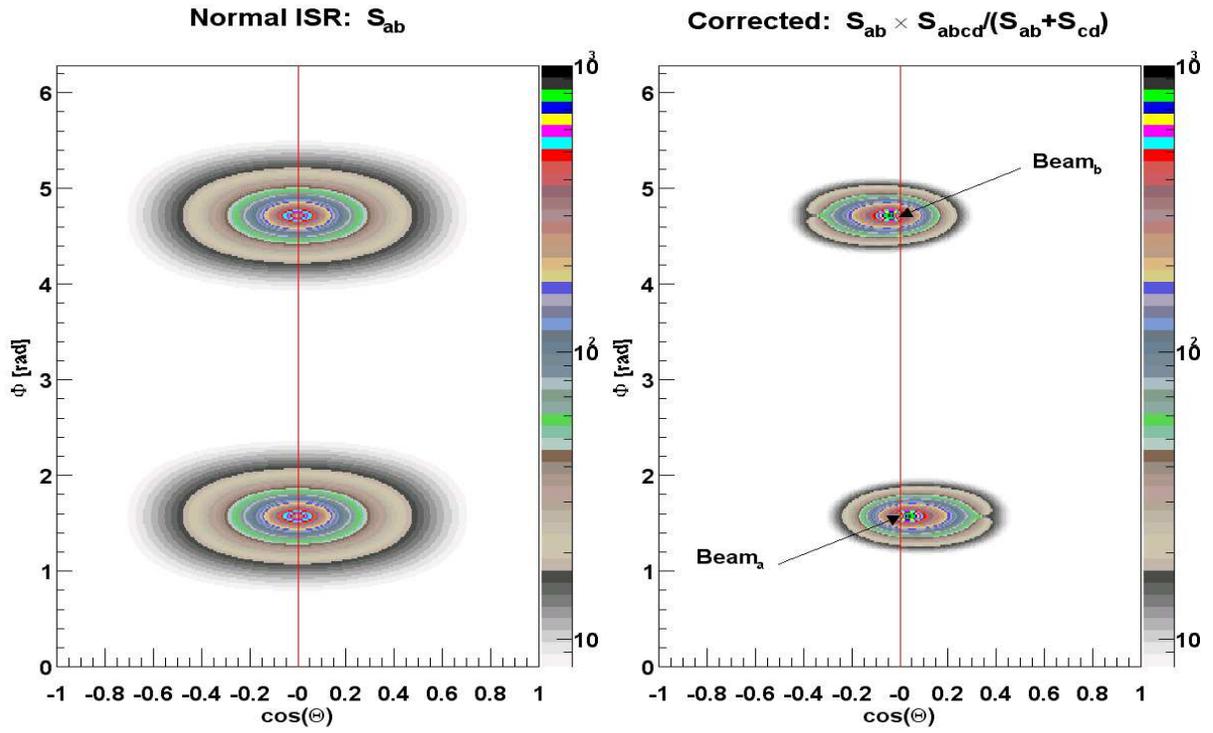,width=160mm,height=100mm} 
\end{center}
\caption{\sf\small
  Photon angular distribution
in ``Mhamha scattering'' 
$\mu^-\mu^+\to \mu^-\mu^+\gamma$ at $\sqrt{s}=5$ GeV and muon scattering
angle of 20$^\circ$ for the case of ISR only.
The difference between left-
  and right-hand side plots shows the effect of the ECS
  correction weight.
}
\label{fig:Mhamha4}
\end{figure}

Having seen in Figs.~\ref{fig:Mhamha1} and \ref{fig:Mhamha2} what the
coherence ECS effect is, let us now ask the important practical question:
Provided that, as in KoralW, we have only ISR, could we modify the ISR
distribution in such a way that we get the same sharpening of the photon angular
distribution, as in the real world of the coherent $\tilde{S}_{abcd}$?
Our proposal for the correcting weight is the following:
\begin{equation}
  W_{\rm ECS}(k) = \frac{\tilde{S}_{abcd}(k)}{\tilde{S}_{ab}(k)+\tilde{S}_{cd}(k)},
  \label{eq:w_ecs}
\end{equation}
and we are going to show that it does what we want.
Let us examine the corrected ISR distribution
\begin{equation}
  \tilde{S}_{ab}(k) W_{\rm ECS}(k).
\end{equation}
First, we notice that $W_{\rm ECS}(k)\simeq 1$ (apart from the
unimportant helicity zero) 
close to the angular position of the outgoing particles,
so it does not try to re-install the FSR,
and the corresponding MC weight should not have an inconvenient long tail.
Then, what is most important, $W_{\rm ECS}$ really cuts off very strongly photon
emission at angles greater than
the scattering angle $\theta_{ac}$ or $\theta_{bd}$.
This is seen very clearly in Fig.~\ref{fig:Mhamha4}, where the ``angular size''
of the corrected ISR emission pattern
shrinks by a factor of $\sim 3$ with respect to the original one.
This is clearly visible even for our moderately small scattering angle $\theta_{ac}=20^\circ$.
In the real KoralW application we shall, of course,  apply the product of the
$W_{\rm ECS}(k)$ over all ISR photons.
The overall normalization will also be corrected by means of correcting the exponential
virtual+real form factor, which also compensates for the IR divergence
of the average ECS weight $\langle \prod_i W_{\rm ECS}(k_i) \rangle$.

How does one justify the fact that we apply a correcting weight, which
includes ISR$\otimes$FSR interference, without generating FSR in the MC?
A quite general answer is that the above can be understood
as a procedure of integrating analytically over the real FSR photons and
including the result of the integration in the normalization%
\footnote{The FSR photons can be added later on; however, care must be taken
  to get the ECS for the FSR in a similar way as for ISR.}.
Such a procedure can be very close to experimental reality, provided
the FSR photons are combined with the outgoing charged particles.
From the theoretical point of view the above procedure is quite common
in the world of the LL approximation.
For example, a similar solution is used to install interference between 
different branches of FSR in the PHOTOS Monte Carlo \cite{photos:1994}.
As we shall see later on, our procedure not only makes sense in the LL approximation,
but also features the precise IR cancellations
and has the expected LL behaviour.

\section{ECS Correction weight for the $4f$ process}
\label{sec:ECSwt}

We have already explained what the ECS effect is, and sketched how to
introduce
it in an approximate way in the ISR Monte Carlo, with the help of the 
correcting weight
$W_{\rm ECS}$.
All that was done for the $2f$ final state in a rather {\em qualitative} way.
In the following we define the analogous correcting procedure for the
four-fermion ($4f$) 
final state; we shall also present several  {\em quantitative} tests,
see Section~\ref{sec:numerics}.

\subsection{Real emission part of $W_{\rm ECS}$ }

Let us start by defining the real emission part of the $W_{\rm ECS}$
for the  $4f$ process
$e^-e^+\to f_c(p_c) +\bar{f}_d(p_d) +f_e(p_e) +\bar{f}_f(p_f) $.
For this process the complete soft photon emission factor reads
\begin{equation}
\tilde{S}(k)=  \frac{1}{2} \sum_{ {i,j=a,b,...,f \atop i\neq j } }
\tilde{S}_{ij}(k),\qquad 
\tilde{S}_{ij}(k)=
   Z_{ij}\frac{\alpha}{4\pi^2} \left(\frac{p_i}{kp_i} -
\frac{p_j}{kp_j}\right)^2, 
\label{eq:stilde}
\end{equation}
where $Z_{ij}= Z_{i}Z_{j}\theta_i\theta_j$, $Z_i$ is the sign of the
$i$-th charge, and $\theta_i = + \,(-)$ if particle $i$ is outgoing
(incoming). 
The natural extension of the weight of Eq.~(\ref{eq:w_ecs})
would be
\begin{equation}
  W_{\rm ECS}^{real}
   =\frac{\tilde{S}(k)}{\tilde{S}_{ab}(k)+\tilde{S}_{cdef}(k)}, 
\label{eq:wtreal0}
\end{equation}
where $\tilde{S}_{cdef}$ is defined as in Eq.~(\ref{eq:stilde}),
restricting summation to the FSR part, $i,j=c,d,e,f$.
The above weight could be implemented in the MC without much problem;
however, it would complicate the construction of the accompanying
normalization weight 
$W_{\rm ECS}^{norm}$ described in the next section.

We have noticed, however, that since we are working in the LL framework,
we can simplify 
the weight of Eq.~(\ref{eq:wtreal0}), and effectively replace it with
the variant of the 
$2f$ weight of Eq.~(\ref{eq:w_ecs}).
How is this possible?
One has to keep in mind that our aim is to deal with the situation in
which one of the final
state particles (electrons) is close to the beam or two particles
(electrons/positrons) 
are close to the beams.
Let us call the above two situations (following a long established terminology)
singly- and doubly-peripheral scattering.

In the singly-peripheral (SP) situation we expect ECS in only one hemisphere.
Photons will be emitted inside a narrowly collimated $t$-channel dipole
close to the beam;
there will be {\em a gap in the photon angular distribution} 
extending from the dipole down
to the nearest particle in the ``central region''%
\footnote{By central region we understand angles greater than, say, $10^\circ$
  from both beams.}
out of the remaining three final particles.
In the doubly-peripheral (DP) configuration, we shall have two narrowly
collimated 
$t$-channel dipoles close to the beams and two gaps down to the central region.
In fact the two gaps will join together in the central region.
This gap structure will be clearly seen in the numerical results in 
Section~\ref{sec:numerics}, where we shall plot the photon distribution in the
rapidity 
variable
$y=-\ln\tan(\vartheta_\gamma/2)$.

Now, if our main aim is to reproduce one or two of the ECS gaps in ISR
radiation, 
and if we restrict ourselves to the LL approximation, then it does not
matter how many 
particles there are in the central region.
What {\em really} matters is how many (one for SP and two for DP case)
particles close to the beams there are, and at which angles those
closest to the beams (members of $t$-channel dipoles) are.
Consequently, in the LL approximation,
for the purpose of reproducing ECS,
it is perfectly safe to replace the
four final particles by just two (especially that there are only two beams).
In the DP case, the choice is clear: we take those two particles
closest to the beams and properly match the beam charges 
(form together with the beams $t$-channel dipoles),
ignoring the other two (charge conservation is of course respected).
In the SP case one charge is again the member of the $t$-channel dipole and another
charge we place anywhere in the central regions;
it may be tied up with one of the three remaining particles or not 
-- it does not matter.

The two final ``effective final state particles'' entering the ECS weight we shall
denote $C$ and $D$, and the real-emission part of our ECS weight we define as follows
\begin{equation}
 W_{\rm ECS}^{real} = \prod_{i} w^{R}(k_i), \;\;\;
 w^{R}(k)=\frac{
    \tilde{S}_{ab}(k) +\tilde{S}_{CD}(k) +\tilde{S}_{aC}(k) +\tilde{S}_{bD}(k) 
   +\tilde{S}_{aD}(k) +\tilde{S}_{bC}(k) }
 { \tilde{S}_{ab}(k) +\tilde{S}_{CD}(k) }.
\label{waga-real}
\end{equation}
The above formula, apart from notation, is identical to that of 
Eq.~(\ref{eq:w_ecs}).
Also, in the numerator we have used the algebraic identity of Ref.~\cite{yfs:1961}
in order to decompose $\tilde{S}_{abCD}$ into six dipole-type contributions.

In practice, the final state fermions $C$ and $D$ in Eq.~(\ref{waga-real})
are determined by the kinematics of the final state (FS).
In the case of the so-called ``single-$W$'' $(1W)$ final state, 
in which we are mainly
interested here, we check whether one final state electron (positron)
is lost in the beam pipe and the decay products of the $1W$ are visible
outside the beam pipe.
In such a case we identify the in-beam-pipe particle as one of the effective
particles $C$ or $D$ (according to its charge).
As for the other one, there is freedom of its choice, as explained above.
Since all other FS fermions are at large angles (central region)
it should not matter which one we pick.
In fact we do something even more primitive: we assign the second
remaining $C$ or $D$ 
as directed perpendicularly to the beam pipe.
In the second, DP case of the $e^-e^+ f\bar{f}$ final state with
both $e^-e^+$ lost in the beam pipe, we assign $C$ and $D$ to $e^-$ and $e^+$, 
of course.

Let us summarize the main points on the weight of Eq.~(\ref{waga-real}),
as implemented in KoralW:
\begin{itemize}
\item
  The only purpose of the weight $W_{\rm ECS}^{real}$ is to restore the 
  ECS effect due to ISR$\otimes$FSR interference.
\item
  We do not aim at re-creating the FSR. This would be formally possible with
  a similar weight; however, it would lead to an awful weight
  distribution and  a non-convergent
  MC calculation.
\item
  We get $W_{\rm ECS}^{real}\to 1$ for photons collinear with the FS effective fermions
  $C$ and $D$. This ensures a very good weight distribution.
\item
  The FSR can be treated separately, either inclusively (calorimetric acceptance)
  or exclusively, generated with the help of PHOTOS%
  \footnote{Care has to be taken to implement ECS for FSR, if necessary.}.
\end{itemize}

Finally, we notice that in the ECS weight we may insert massless four-momenta
without any problem
-- in practice we shall use a massless limit of Eq.~(\ref{eq:stilde}):
\begin{equation}
  \tilde{S}_{ij}(k)\to\tilde{S}_{ij}(k)= Z_{ij}\frac{\alpha}{4\pi^2}
  \frac{-2p_ip_j}{(kp_i)(kp_j)}.
\label{stilde0}
\end{equation}

\subsection{Virtual+soft correction to normalization}
\label{section:v+s}

The average of the real emission weight $\langle W_{\rm ECS}^{real} \rangle$
taken alone is infrared (IR)-divergent as $\sim \ln\epsilon$,
where $\epsilon$ defines the infrared cut $E_\gamma> \epsilon E_{beam}$
on the photon momentum in the CMS.
In the YFS exponentiation of ISR in KoralW, based on Ref.~\cite{yfs2:1990},
the IR cancellations occur {\em numerically} between the real 
soft-photon factors
$\prod_i \tilde{S}(k_i)$
and the YFS form factor $F_{YFS}=\exp(2\alpha B+\tilde{B}(\epsilon))$
(contrary to typical parton-shower MCs, where they are built-in features
of the MC algorithm).
Since it is well known how these IR cancellations do occur,
see refs.~\cite{yfs2:1990,Jadach:2000ir},
it is therefore also possible to calculate the IR violation of
the $\langle W_{\rm ECS}^{real} \rangle$, and to correct for the above 
divergence {\em precisely}.
At the same time, we shall also correct for the LLST in the virtual
part of the YFS form factor and the non-infrared LL corrections to the
${\cal O}(\alpha)$.
As a result, the corrected total cross section will become again independent
of the IR parameter $\epsilon$, as the original one.
This we shall check numerically.
Again, all of this procedure should be regarded as a ``shortcut''
with respect to the full YFS exponentiation with ISR+FSR including 
ISR$\otimes$FSR interference, 
keeping in mind that our precision tag is limited to 2\%,
and that simplicity of the method is also our high priority.

The correction to the overall normalization,
which corrects for the IR divergence exactly, 
and introduces LLST in the virtual LL correction,
reads as follows:
\begin{equation}
\begin{split}
& W_{ECS}^{norm} = \exp \left(\frac{3}{4}(\bar\gamma_t -\gamma_s)\right) 
                   \exp \big(\Delta U(\epsilon)\big) 
\\
& \Delta U(\epsilon)= U(\epsilon) -U_R(\epsilon), \;\;
U(\epsilon)= \int\limits_{\epsilon \sqrt{s}/2}^{\sqrt{s}} \frac{d^3k}{k^0}
                      \tilde{S}_{ab}(k), \;\;
U_R(\epsilon)= \int\limits_{\epsilon \sqrt{s}/2}^{\sqrt{s}} \frac{d^3k}{k^0}
                      \tilde{S}_{ab}(k)w^R(k).
\label{waga-vs}
\end{split}
\end{equation}
The factor $\exp(\Delta U)$ cancels {\em exactly} the $\epsilon$-dependence
and compensates {\em approximately} for the normalization change due
to the $\langle W_{\rm ECS}^{real} \rangle$ weight. The proof and details
of construction of $\exp(\Delta U)$ are given in Appendix \ref{app1}.

The factor $\exp((3/4)(\bar\gamma_t -\gamma_s))$ provides the LLST in
the virtual part of the form factor.  The $\gamma_s$ is defined as
$\gamma_s = 2(\alpha/\pi)(\log(s/m^2_{e})-1)$. The scale of the
$\bar\gamma_t$ depends on the number of final state particles that
effectively contribute to the interference weight (\ref{waga-real}).
When only one line is modified, say $e^-$, we have 
$\bar\gamma_t = (1/2) (\gamma_s +\gamma_{t^-})$, where $\gamma_{t^-} =
2(\alpha/\pi)(\log({|t^-|}/m^2_{e})-1)$ and $t^{-}$ denotes the square
of the four-momentum transfer
from the $e^-$ line. 
When both $e^-$ and $e^+$ lines are modified, we get 
$\bar\gamma_t = (1/2) (\gamma_{t^-} +\gamma_{t^+})$.

The triple integral of Eq.~(\ref{waga-vs}) has to be computed for every
generated event. It is therefore of high importance to be able to do it
very fast, so that the Monte Carlo generation is not slowed down too
much.
We have not attempted to compute it completely in an analytic way. 
Instead, we perform two out of three integrations analytically and the
remaining one numerically. Such a procedure proved to be fast enough.
We show in Appendix \ref{app2} how to reduce the integral in 
Eq.~(\ref{waga-vs}) to a simple one-dimensional integral.

\section{Running QED Coupling Constant}

In the practical applications, a more precise prediction for the $1W$-type
processes requires also the inclusion of the effect of the running QED
coupling constant in the $e^-\, (e^+)$ vertex from the value at $M_W^2$ scale
to the actual small transfer value $t^-\, (t^+)$.  
We include this effect in the form of an overall factor 
that multiplies the whole matrix element squared. 
It is activated only for the electron or positron
line (in the Feynman diagram) for which the ECS corrective weight is
actually applied. 
In the SP case of one $e^\pm$ line, the correcting weight reads:
\begin{equation}
  W_{Run} =W_{Run}^\pm = \left( \frac{\alpha(t^\pm)}{\alpha_{G_\mu}} \right)^2,
\end{equation}
where $\alpha_{G_\mu}$ is the
value of coupling constant in the G$_\mu$ scheme%
\footnote{
  If any scheme other than G$_\mu$ is used in { KoralW}, the program
  will report a conflict and stop at this point.}. 
For the DP case (two lines) we put
\begin{equation}
  W_{Run} = W_{Run}^- W_{Run}^+.
\end{equation}
The precision of such a naive solution is, in most cases 
(including leptonic final states), better than 2\%, 
and for semileptonic final states even better than 1\%,
as discussed at length in Ref.\ \cite{Passarino:2000mt}.

\section{Numerical results}
\label{sec:numerics}

In this section we check very carefully that our
ECS weight
\begin{equation}
  W_{ECS} = W_{ECS}^{norm}\; W_{ECS}^{real}\; W_{Run}
\end{equation}
introduces the ECS effects in the photon angular distribution, the LLST
in the total energy loss due to ISR photons,
and corrects the QED coupling constant for the hard process.

\subsection{General consistency tests}
In this subsection we present some numerical results that demonstrate the
action of the weight emulating the ECS effect.

The first test, necessary for consistency, shows that the cross-section
is independent of the dummy IR cut-off $\epsilon$ of Eq.\
(\ref{waga-vs}). In Table 1 we show the values of the total 
cross sections for
the $e\bar\nu_e u\bar d$ channel for a
few values of the dummy IR cut-off $\epsilon_{}$. The energy is set to
190 GeV,
and the cut-off on the maximal angle of the scattered
electron is $2.5^\circ$ w.r.t.\ the electron beam in the
effective CMS frame of outgoing particles.
One can see that, within the statistical errors, there is indeed
no dependence on the $\epsilon$. 
\begin{table}[h]
\centering
\begin{small}
\begin{tabular}{|c|r|r|r|r|r|r|r|}
\hline
$\epsilon_{}$   &    
    $10^{-3}$&   $10^{-4}$ & $10^{-5}$ &    $10^{-6}$&    $10^{-8}$&    $10^{-9}$&   $10^{-10}$  
\\ 
\hline 
$\sigma$     &      
      0.09773&      0.09757&      0.09755&      0.09752&      0.09762&      0.09737&      0.09759  
\\ 
    {[pb]}   &
$\pm$ 0.00020&$\pm$ 0.00022&$\pm$ 0.00010&$\pm$ 0.00019&$\pm$ 0.00019&$\pm$ 0.00020&$\pm$ 0.00011
\\
\hline
\end{tabular}
\end{small}
\caption{\small\sf Value of the total cross section for the 
  $e\bar\nu_e u\bar d$
  final state, for various values of the dummy
  IR cut-off $\epsilon_{IR}$.
}
\end{table}

\begin{figure}[!ht]
\setlength{\unitlength}{0.1mm}
\hskip 1cm
\epsfig{file=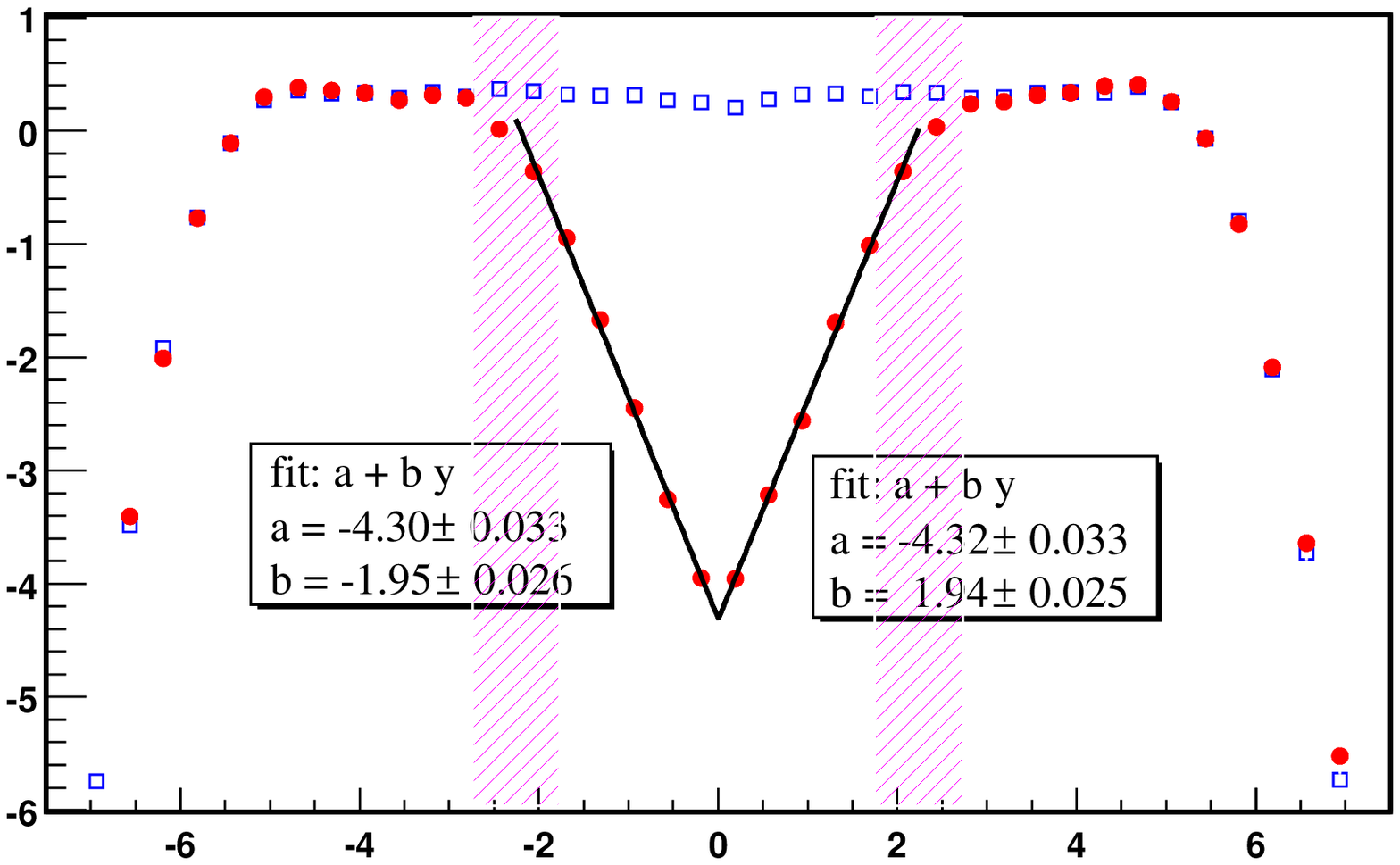,width=160mm,height=100mm}
\caption{\small\sf
  $\log_{10}$ of 
  $d\sigma/d\log_{10}\tan(\theta_{\gamma}/2)$ 
  with (red dots) and without (blue open squares) the
  ECS correction, arbitrary units. In boxes the values of fits are
  shown.
}
\begin{picture}(1600,0)
\put(1000,190){\large $y=-\log_{10}\tan(\theta_{\gamma}/2)$}
\put(0,960){\large $\log_{10}\frac{d\sigma}{dy}$}
\put(0,880){\large  [arb.~units]}
\end{picture}
\label{fig:Fig1}
\end{figure}

Having checked the self-consistency of the emulation, we can now proceed
to demonstrate the main result of this work -- the ECS suppression
of the transverse radiation. In Fig.~\ref{fig:Fig1} we show 
(in the doubly-logarithmic scale) the differential distributions
$d\sigma/dy$
with (red dots) and without (blue open squares) the correction weight for the
ECS effect for
the $e^+e^- s\bar s$ channel. The angles of scattered
electron and positron are set to be between 2 and 0.2$^\circ$ 
w.r.t.\ the beam 
line (the double
``ice-cream-cone'' configuration around the beam line) in the laboratory
frame. 
The distribution is obtained by summing over {\em all} photons of a
given event and is normalized to arbitrary units. 
The $y=-\log_{10}\tan(\theta_{\gamma}/2)$ is proportional to 
the (pseudo)rapidity of the photon.

The scattering cone of electron (positron) corresponds, in the $y$ 
scale, to the values from $1.76$ to $2.76$ ($-1.76$ to $-2.76$) and are
marked in the plot with blue shadow bands. In the Figure
one can clearly see the ECS suppression between these
bands. The spectrum for photon emission outside the area (i.e.~inside
the $e^-$ and $e^+$
cones) remains flat and unchanged. Between the bands (i.e.~outside the
$e^-$ and $e^+$ cones) the suppression is clearly visible. We have
fitted the suppressed spectrum with straight lines. The
values of the fitted slopes are shown in the figure to be in good
agreement 
with the values of $\pm 2$, which means that we see the 
$1/ \theta_{\gamma}^2$
suppression of the photons beyond the dipole angular size, naively
expected from the ``multipole expansion''.

\begin{figure}[!ht]
\setlength{\unitlength}{0.1mm}
\hskip 1cm
\epsfig{file=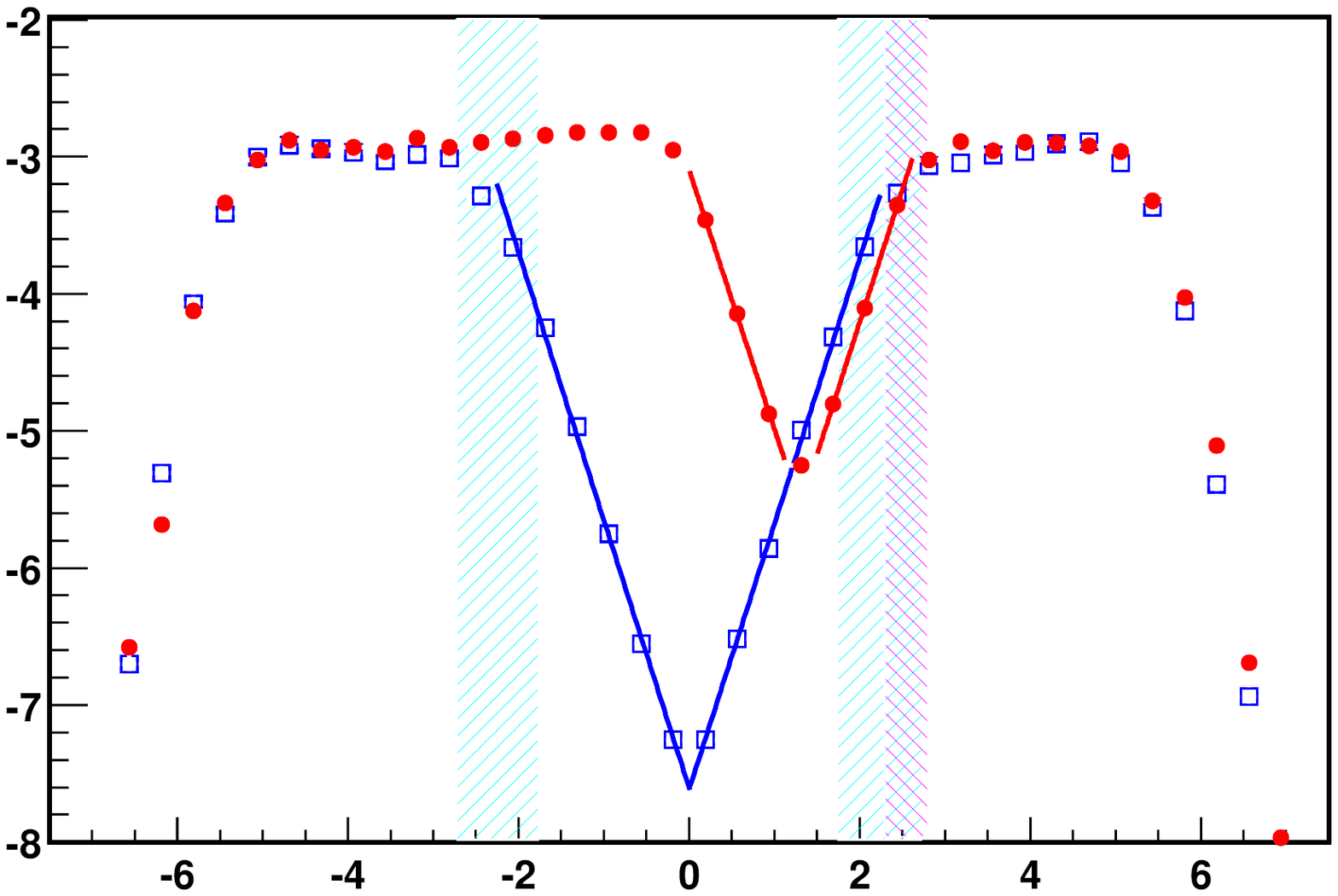,width=160mm,height=100mm}
\caption{\small\sf
  $\log_{10}$ of 
  $d\sigma/d\log_{10}\tan(\theta_{\gamma}/2)$ 
with the
ECS correction for $e\nu_e u\bar d$ (red dots) and $e^+e^-s\bar s$ (blue
open squares), arbitrary units.
}
\begin{picture}(1600,0)
\put(1000,190){\large $y=-\log_{10}\tan(\theta_{\gamma}/2)$}
\put(0,960){\large $\log_{10}\frac{d\sigma}{dy}$}
\put(0,880){\large  [arb.~units]}
\end{picture}
\label{fig:Fig2}
\end{figure}

In the next step we look into the $1W$-type final state of
$e\bar\nu_e u\bar d$. This time, there is only one electron in the final
state, so we expect the suppression area to be asymmetric -- only in the
forward direction. The result of the simulation is shown in
Fig.~\ref{fig:Fig2}. 
In blue (open squares) we show the same curve as in Fig.~\ref{fig:Fig1}, 
i.e.~the $e^-e^+s\bar s$ case (appropriately renormalized), 
whereas in red (dots) we present the new 
$e\bar\nu_e u\bar d$ result. Shadowed bands again visualize the location
of the electron (positron) ``ice-cream-cones''. For the  $e\bar\nu_e
u\bar d$ case the area corresponds to the ``ice-cream-cone'' between 0.2 and 
0.5$^\circ$. The other end of the $e\bar\nu_e u\bar d$ suppression area is
located at zero ($90^\circ$), where the fictitious particle is located
and leaves the radiation in the backward hemisphere unchanged.
Note that the exact location of this fictitious particle (in the area of
reasonably large angles)
is irrelevant anyway, as the effect depends on it logarithmically.
The fitted lines can be used as before to confirm the correctness of the
suppression factor (slope).

\begin{figure}[!ht]
\setlength{\unitlength}{0.1mm}
\hskip 1cm
\epsfig{file=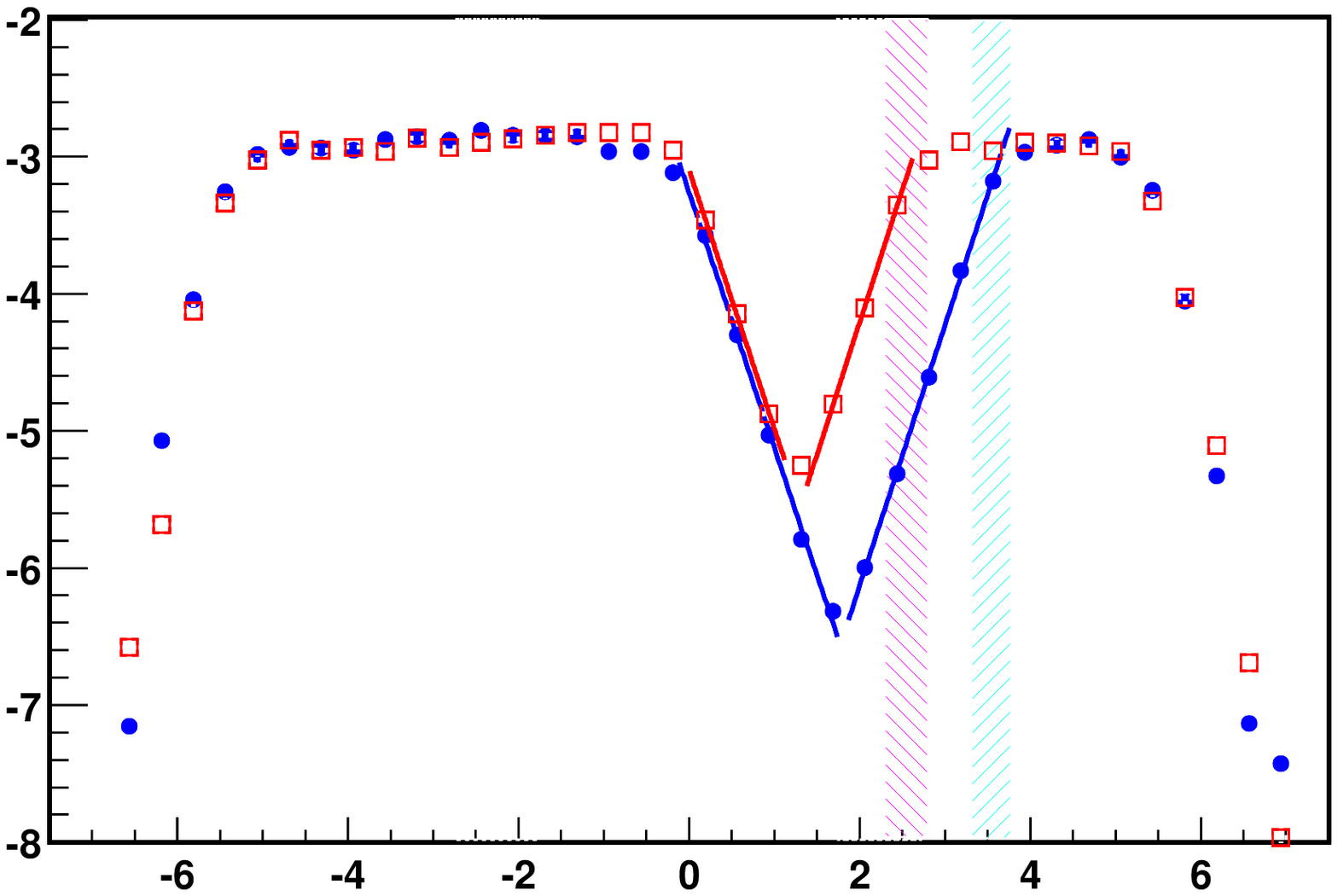,width=160mm,height=100mm}
\caption{\small\sf
  $\log_{10}$ of 
  $d\sigma/d\log_{10}\tan(\theta_{\gamma}/2)$ 
with the
ECS correction for $e\nu_e u\bar d$ for two values of cuts for 
electron opening angle, arbitrary units.
}
\begin{picture}(1600,0)
\put(1000,190){\large $y=-\log_{10}\tan(\theta_{\gamma}/2)$}
\put(0,960){\large $\log_{10}\frac{d\sigma}{dy}$}
\put(0,880){\large  [arb.~units]}
\end{picture}
\label{fig:Fig3}
\end{figure}

Finally, in Fig.~\ref{fig:Fig3} we show the $e\bar\nu_e u\bar d$ process
for two 
different values of the half-opening angle of the electron
``ice-cream-cones'': 0.2 to 0.5$^\circ$ and 0.02 to 0.05$^\circ$. One can
see that the left (backward) edge of the suppression area is, as
expected,  identical in the two 
cases, whereas the right (forward) edge follows the electron
opening angle (the range of this angle is marked with shadowed bands).

Another characteristic distribution is
 the $d\sigma/d\log_{10}v$ distribution, with the definition
 $v=1-s'/s$. As this distribution is sensitive only to the value of the
 transfer in the hard scattering process, we have defined the
 acceptance cuts for this exercise to be $ -9 > \log (t/s) > -11$. 
In the limit of small $v$ (soft limit) we know
from YFS exponentiation that%
\footnote{
The YFS exponentiation alone leads only to the $\exp(\gamma/4)$
factor. The additional $\exp(\gamma/2)$ comes from 
leading-logarithmic considerations.
In the YFS scheme this factor is supplied order by order by
perturbative non-infrared corrections.}
$d\sigma/dv = \exp(3/4\gamma) F(\gamma) \gamma v^{\gamma-1} \sigma((1-v)s)$ 
where 
$F(\gamma) = \exp(-\gamma C_{Euler})/\Gamma(1+\gamma) = 1 +{\cal O}(\gamma^2)$.
The $\gamma$ for the case of pure $s$-channel ISR is governed by the
scale $s$, i.e.~ 
$\gamma_s = 2\alpha/\pi(\log(s/m^2_{e})-1)$. Inclusion of the ECS
suppression leads to the replacement of the scale $s\to |t|$, i.e.~
$\gamma_t = 2\alpha/\pi(\log(|t|/m^2_{e})-1)$.
In the case of the $1W$-type process ($e\bar \nu_e u \bar d$) this
suppression 
happens only in the electron line, whereas for the positron line we
still retain the scale $s$, so the effective $\bar\gamma_t$ becomes
$\bar\gamma_t = 1/2 (\gamma_s +\gamma_t)$. For the  $e^+e^-$ case, the
suppression happens for both lines and we have $\bar\gamma_t = \gamma_t$.

\begin{figure}[!ht]
\setlength{\unitlength}{0.1mm}
\hskip 1.5cm
\epsfig{file=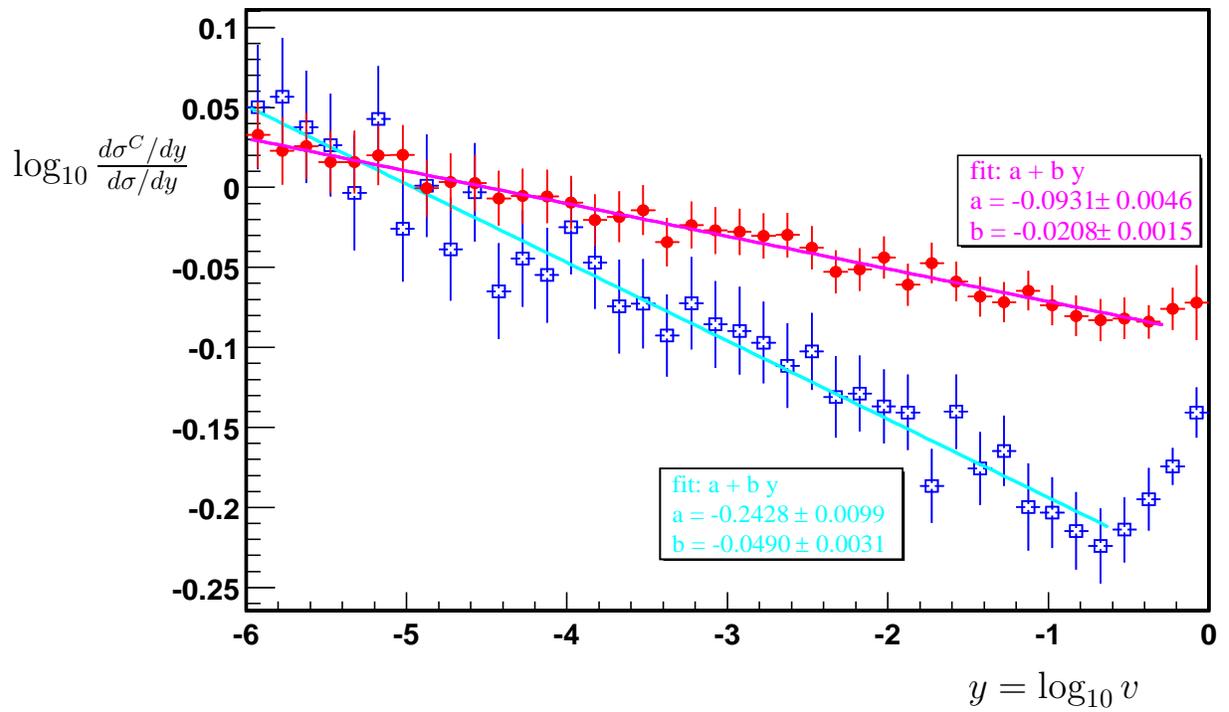,width=160mm,height=100mm}
\caption{\small\sf
$\log_{10}$ of the ratio of $(1/v)(d\sigma/d\log_{10}v)$, $v=1-s'/s$, 
with and
without the ECS correction for $e\bar \nu_e u \bar d$ (red dots)
and $e^+e^-s\bar s$ (blue open squares) final states.
}
\begin{picture}(1600,0)
\put(0,860){\large $\log_{10}\frac{d\sigma^{C}/dy}
                                  {d\sigma/dy}$}
\put(1270,165){\large $y=\log_{10}v$}
\end{picture}
\label{fig:Fig4}
\end{figure}

In Fig.~\ref{fig:Fig4}, we show a ratio of $d\sigma/d\log_{10}v$ distributions 
with and without the ECS correction for the processes  $e^+e^-s\bar
s$ (blue open squares) and for $e\bar \nu_e u \bar d$ (red dots).
For the $e^+e^-s\bar s$ process we applied an additional cut-off of $3.5$
mrad for the minimal angle of $s$ and $\bar s$ quarks with respect 
to the beam line  and
we used the so-called ``extrapolation procedure'' that preserves the
value of smallest transfer (see KoralW manual \cite{koralw:2001} for details). 
As follows from the above discussion, we
expect to see (in doubly-logarithmic scale) straight lines of the form 
$a + b \log_{10}(v)$ where the slope of the line should hence be given by
$b_{th} = \bar\gamma_t -\gamma_s = \alpha/\pi\log(t/s)$ and the free
coefficient by 
$
a_{th} = 
   3/4 (\bar\gamma_t-\gamma_s)\log_{10}(e) 
 +                          \log_{10}(\bar\gamma_t/\gamma_s)
$.
In our exercise the
 average value of the $\log (t/s)$ as calculated in the simulation
 is  $\log (t_0/s)=-9.86$. 
This leads, at $t=t_0$, to the expected theoretical values of slopes of
$b_{th}=-0.023$ and $a_{th}= -0.104$ for $e\bar \nu_e u \bar d$ and 
$b_{th}=-0.046$ and $a_{th}= -0.241$ for $e^+e^-s\bar s$. 
The corresponding results of the fits of 
 the actual Monte
Carlo simulations to straight lines, 
shown in the insets in Fig.~\ref{fig:Fig4}, are 
$b_{th}=-0.021\pm 0.002$ and $a_{th}= -0.093\pm 0.005$ 
for $e\bar \nu_e u \bar d$ and 
$b_{th}=-0.049\pm 0.003$ and $a_{th}= -0.24\pm 0.01$ for $e^+e^-s\bar s$. 
We see an agreement at the level of one standard deviation for the 
$e^+e^-$ case and two standard deviations for the case of one electron.
The latter discrepancy can be a signal that our naive expectation for the slope
is good only to a few per cent of its value.
The value of the free coefficient is in principle 
sensitive to the overall normalization factor ($3/4$).
It is however numerically dominated by a much larger 
$\log_{10}(\bar\gamma_t/\gamma_s)$ term; the actual factor of $3/4$
contributes below 10\%, i.e.~at the level of accuracy of the whole 
comparison, and is, therefore, inconclusive.

\subsection{Predictions for the $1W$ process}
In this subsection we present a few results of the influence of 
the ECS effect in { KoralW} on the $1W$-type process 
$e^+e^- \to e^-\bar\nu_e u\bar d$. We use the following cut-offs 
(similar to Ref.\ \cite{4f-LEP2YR:2000}):
\begin{itemize}
\item
electron angular acceptance: $\cos\theta_e \geq 0.997$,
\item
quark--antiquark invariant mass: $M_{q\bar q} \geq 45$ GeV,
\end{itemize}
and the following setup:
$G_\mu$ scheme; 
fixed $W$ and $Z$ widths;
normal (non-screened) Coulomb correction;
naive QCD correction;
extrapolation procedure that fixes the smallest transfer ({\tt KeyISR=3});
$W$ branching ratios with mixing and naive QCD correction
 calculated in IBA from the CKM matrix \cite{PDG:2000};
G$_\mu=1.16639\times 10^{-5}$,
$M_Z=91.1882$ GeV, $M_W=80.419$ GeV, $\Gamma_Z=2.4952$ GeV,
$\alpha_S=0.1185$.
\begin{table}[h]
\centering
\begin{small}
\begin{tabular}{|c|r|r|r|r|}
\hline
E [GeV] & $\phantom{\Bigl(}\sigma^{C+R}$ [fb]   
        & $1-\sigma^{R}/\sigma^{C+R}$ [\%]
        & $1-\sigma^{C}/\sigma^{C+R}$ [\%]
        & $1-\sigma^{ }/\sigma^{C+R}$ [\%]
\\
\hline
190 & $87.11 \pm 0.22$ 
    & $ 5.65 \pm 0.15$
    & $-5.16 \pm 0.02$
    & $ 0.81 \pm 0.14$
\\
\hline
200 & $103.60 \pm 0.26$ 
    & $  5.49 \pm 0.15$ 
    & $ -5.13 \pm 0.02$
    & $  0.67 \pm 0.15$
\\
\hline
500 & $807.56 \pm 2.74$ 
    & $  4.92 \pm 0.21$
    & $ -4.68 \pm 0.02$
    & $  0.51 \pm 0.21$
\\
\hline
\end{tabular}
\end{small}
\caption{\small\sf The total cross-section with ECS effect (C) and running
  QED coupling (R). The cut-offs and parameter setup are as described in
  the text. The relative size for the ECS effect and the running
  QED coupling effect are also shown.
}
\end{table}

In Table 2 we present the values of cross sections with the ECS
effect (marked with C) and running of the QED coupling (marked with R) 
for a few energies in
the LEP2 and future LC energy range. In the respective 
columns we show separately 
the changes 
due to the ECS effect (third col.), the running of the
QED coupling (fourth col.) and both corrections together (fifth
col.). All changes are given in per cent with respect to the corrected cross
section. One can see that the reduction of ISR due to ECS
suppression increases the cross
section for all
energies, consistently with the overall effect of ISR. The running of
the QED coupling decreases the cross section by roughly the same amount
regardless of the energy.
The relative shifts for LEP2 energies presented in Table 2 
are in a qualitative agreement with the results of Ref.\ \cite{Montagna:2000ev}.

\subsection{Precision of the $1W$ cross sections}

Finally, we have to address the question of the physical precision of 
the above $1W$ cross sections. There are two components of the error 
-- approximations in the ECS treatment and approximations 
related to the running QED coupling and higher order EW effects. 
\begin{itemize}
\item
The approximations in the treatment of ECS are of the genuine
non-leading type,
i.e.\ $\alpha/\pi$ with possible enhancement factors (e.g.\
$\pi^2$). 
Therefore we estimate this error to be below 2\%.
\item 
Following Ref.\ \cite{Passarino:2000mt},
we estimate the approximation due to the naive treatment of running the QED 
coupling to be at most 2\%, dominated by the leptonic channel contribution. 
In some specific cases, such as the semileptonic  
final states, this error can be lowered to 1\% 
(see \cite{Passarino:2000mt} for details).
\item 
The other missing EW effects we estimate at 1\%, again
following Ref.\ \cite{Passarino:2000mt}.
\end{itemize}

To summarize, in the new version 1.53 of KoralW,
upon adding the above contributions in quadrature, we obtain
an overall 
precison tag of the single-$W$ cross sections of 3\%. 
This number can be reduced for 
some specific final state configurations, because of a smaller error 
contribution from the running QED coupling.
The precision of the ECS implementation alone is of the order of 2\%.
Such overall theoretical precision of  3\% for the single-$W$ cross
sections lies well below the expected
final LEP2 experimental accuracy of $\sim 7$\% \cite{ICHEP2002}.

\section{Summary}

We have shown that it is possible to improve the standard ISR calculation
in such a way that it takes into account electric charge screening
and LL scale transformation for singly- and doubly-peripheral configuration
in the production of the $4f$ final states, with electron and/or positron
in the beam pipe.
This is done using a relatively simple and well-behaving MC weight.
The method does not necessitate an explicit inclusion of the FSR.
Although the method is not the exact implementation  of the YFS 
exclusive exponentiation for the ISR+FSR, it is nevertheless closely
modelled upon it.
The QED running coupling constant for the hard process is corrected
at the same time to the correct $t$-channel scale.
We also present a number of numerical consistency checks and a sample
prediction for the $1W$ total cross section.
The MC program KoralW (in its new version 1.53) in which this method is 
implemented is
readily available for any interested user.
It will be especially useful for the final LEP2 data analysis.

\section*{Acknowledgements}
\noindent
We would like to thank A. Valassi for stimulating discussions. 
We warmly acknowledge the kind support of the CERN TH and EP divisions.

\appendix
\section{Virtual+soft normalization weight}
\label{app1}
In this appendix we discuss the construction of the {\em ad hoc} 
normalization weight of Eq.\ (\ref{waga-vs}) and show that it ensures 
independence of the cross section from 
the dummy infrared cut-off $\epsilon$. 

Before going into details, let us stress again that we introduced the ECS
ansatz {\em by hand} into the pure ISR-type MC algorithm in order to
cure some low-angle-emission pathologies. Therefore it is {\em not} the
purpose of this
appendix to rigorously derive the ECS ansatz from ``basic
principles''. Such a rigorous introduction of the ECS requires an entire
reformulation of the MC algorithm, from ISR-type to ISR$\otimes$FSR-type, based
on the complete YFS theory and has already been done in other MC programs,
such as BHWIDE~\cite{bhwide:1997} or KKMC~\cite{Jadach:1999vf,Jadach:2000ir}.

We will use the Mellin transform representation of the MC master formula 
as given in Appendix A of Ref.\ \cite{yfs2:1990}. In order to establish
the notation, we will, following 
Ref.\ \cite{yfs2:1990}, briefly recall
the relation of this representation to the standard Monte Carlo form of 
refs.\ \cite{koralw:1995a,koralw:1998,koralw:2001}. Next we will 
show how the real emission weight is added into the master formula and 
what the matching virtual+soft compensation weight must 
look like to cancel the fictitious $\epsilon$ dependence. 

We start from the Mellin transform representation of the master formula 
as given in Eq.\ (A1) of Ref.\ \cite{yfs2:1990} 
adapted to the four-body final state (for details 
we refer to Ref.\ \cite{yfs2:1990})
\begin{equation}
\label{mastermel}
\begin{split}
 & \sigma =
  \lint {d^4x\over (2\pi)^4} \;
  \lint \prod_{i=1}^4 {d^3q_i\over q_i^0} \;
\exp\bigg[ix\bigg(p_1+p_2-\sum_{i=1}^4 q_i \bigg) +D\bigg]
  \exp\big(  2\alpha B
              + 2\alpha \tilde B
      \big)
\\ &
  \Bigg[
         \bbeta^{(3)}_0(p^\Rcal_r,q^\Rcal_s)
      +
         \lint {d^3k\over k^0} e^{-ixk} 
         \bbeta^{(3)}_{1}( p^\Rcal_r, q^\Rcal_s, k)
      +
         \frac{1}{2!}
         \lint {d^3k_1\over k_1^0} {d^3k_2\over k_2^0} e^{-ixk_1-ixk_2} 
         \bbeta^{(3)}_{2}( p^\Rcal_r, q^\Rcal_s, k_1, k_2)
\\ &
      +
         \frac{1}{3!}
         \lint {d^3k_1\over k_1^0} {d^3k_2\over k_2^0} 
               {d^3k_3\over k_3^0} e^{-ixk_1-ixk_2-ixk_3} 
         \bbeta^{(3)}_{3}( p^\Rcal_r, q^\Rcal_s, k_1, k_2, k_3)
  \Bigg],
\end{split}
\end{equation}
where
\begin{equation}
\begin{split}
2\alpha \tilde B &= \lint {d^3k\over k^0} \tilde{S}_{12}(k)
              \theta(\sqrt{s} -k^0)
\\
D&=\lint {d^3k\over k^0} \tilde{S}_{12}(k)
              \bigg(e^{-ixk} -\theta(\sqrt{s} -k^0)\bigg).
\end{split}
\end{equation}
Now the dummy IR cut-off $\epsilon$ is introduced. 
Keeping in mind that  
the $d^4x$ integral provides even stronger cut-off than 
$\theta(\sqrt{s} -k^0)$ and that in the small-$\epsilon$ limit 
$e^{-ixk}\to 1$, one can rearrange the exponents
\begin{equation}
\begin{split}
D+2\alpha \tilde B &= D' + \tilde B(\epsilon)
\\ 
\tilde B(\epsilon) &=\lint {d^3k\over k^0} \tilde{S}_{12}(k)
            \theta(\epsilon{\sqrt{s}\over 2} -k_0), 
\\
D'&=\lint {d^3k\over k^0} \tilde{S}_{12}(k)e^{-ixk}
            \theta(\sqrt{s} -k^0) \theta\Big(k_0 -\epsilon{\sqrt{s}\over 2}\Big).
\end{split}
\end{equation}
After expanding the $D'$ integral and performing $d^4x$ integration 
one obtains 
the master formula in the familiar Monte Carlo form 
\cite{koralw:1995a,koralw:1998,koralw:2001}:
\begin{equation}
\label{masterjw}
\begin{split}
  \sigma =&
  \sum_{n=0}^\infty {1\over n!}
  \lint \prod_{i=1}^4 {d^3q_i\over q_i^0} \;
  \left(\prod_{i=1}^n       {d^3k_i\over k^0_i}
    \tilde{S}_{12}(k_i) \theta\Big(k^0_i-\epsilon\frac{\sqrt{s}}{2}\Big)
   \right)
  \delta^{(4)}\bigg(p_1+p_2-\sum_{i=1}^4 q_i
  -\sum_{i=1}^n k_i \bigg)
\\
  &
  \exp\bigg(  2\alpha B
              + \lint {d^3k\over k^0} \tilde{S}_{12}(k)
              \theta\Big(\epsilon\frac{\sqrt{s}}{2} -k^0\Big)
      \bigg)
  \Bigg[
         \bbeta^{(3)}_0(p^\Rcal_r,q^\Rcal_s)
        +\sum_{i=1}^n
        {\bbeta^{(3)}_{1}( p^\Rcal_r, q^\Rcal_s, k_i) 
           \over \tilde{S}_{12}(k_i)}
\\&
        +\sum_{ i>j }^n
        {\bbeta^{(3)}_{2}( p^\Rcal_r, q^\Rcal_s, k_i, k_j)
           \over \tilde{S}_{12}(k_i)\tilde{S}_{12}(k_j)}
        +\sum_{ i>j>l }^n
        {\bbeta^{(3)}_{3}( p^\Rcal_r, q^\Rcal_s, k_i, k_j, k_l)
           \over \tilde{S}_{12}(k_i) \tilde{S}_{12}(k_j) \tilde{S}_{12}(k_l) }
  \Bigg].
\end{split}
\end{equation}
Reversing the above procedure 
we see that the {\em ad hoc} introduction of the real emission weight 
$W^{real}_{ECS}$ of Eq.\ 
(\ref{waga-real}) into Eq.\ (\ref{masterjw}) amounts to the modification 
of the $D'$ function: 
\begin{equation}
D' \to D'_R=\lint {d^3k\over k^0} \tilde{S}_{12}(k)w^{R}(k)e^{-ixk}
            \theta(\sqrt{s} -k^0) \theta\Big(k_0 -\epsilon\frac{\sqrt{s}}{2}\Big).
\end{equation}
Our goal is now to find the corresponding virtual+soft normalization weight 
that would {\em exactly} compensate for the change of normalization caused in 
the master formula by the replacement $D' \to D'_R$.
It is evident that such a weight should have the form 
\begin{equation}
W^{norm}_{exact}=\exp(D' -D'_R)=\exp\bigg(
\lint {d^3k\over k^0} \tilde{S}_{12}(k)(1-w^{R}(k))e^{-ixk}
            \theta\Big(k_0 -\epsilon\frac{\sqrt{s}}{2}\Big)
                            \bigg).
\end{equation}
Adding and subtracting the function $\Delta U(\epsilon)$ of Eq.\
(\ref{waga-vs}) 
we obtain 
(again in the small $\epsilon$ limit)
\begin{equation}
\begin{split}
\label{waga-vs-exact}
W^{norm}_{exact}&=       
 \exp(\Delta D(\epsilon) +\Delta U(\epsilon))
\\
\Delta D(\epsilon)&= D -D_R,  \;\;
\\
D_R&=\lint {d^3k\over k^0} \tilde{S}_{12}(k)w^{R}(k)
              \Big(e^{-ixk} -\theta(\sqrt{s} -k^0)\Big)
.
\end{split}
\end{equation}
 Let us now summarize the situation. We have constructed the {\em exact} 
compensating weight $W^{norm}_{exact}$. It would exactly restore the 
normalization of the master formula changed by the real emission weight 
$W^{real}_{ECS}$. At the same time, however, this weight would ruin 
the entire MC algorithm. Because of its $x$-dependence through the
$\Delta D(\epsilon)$ 
function, the series of integrals over $d^4x$ leading to 
four-momenta-conserving 
delta functions could not be performed now! Therefore we have to modify the 
compensating weight. We have to do it in such a way that:
(1) the master formula remains independent of dummy $\epsilon$  
parameter and (2) the normalization of the master formula remains 
as close as possible to the original normalization of 
Eq.\ (\ref{mastermel}).
To fulfil these requirements we observe that the whole dependence on 
$\epsilon$ is contained 
in the $\Delta U(\epsilon)$ integral. We will not modify it and 
the condition 
(1) will be fulfilled. 
The trouble-making $\Delta D(\epsilon)$ function is up to the small term 
$(1-w^{R}(k))$
identical to the original $D$ function. As is known \cite{yfs:1961}, 
the role of the $D$
function is to compensate for the four-momentum non-conservation in the 
case of emission of multiple high energy photons. 
Numerically it is a very small
correction, of the ${\cal O}(\alpha^2)$ LL type. Therefore within our accuracy
we can drop this whole term and the condition (2) will be fulfilled 
within an ${\cal O}(\gamma^2)$ accuracy. Upon including the LLST 
correction factor $\exp \big((3/4)(\bar\gamma_t -\gamma_s)\big)$, 
as described in Section \ref{section:v+s}, we arrive at the  
$W_{ECS}^{norm}$ of Eq.\ (\ref{waga-vs}).

\section{Analytical integration}
\label{app2}
We show here how to calculate analytically the 
$\Delta U(\epsilon)$ function of Eq.~(\ref {waga-vs}).
In the polar variables the $dk^0$ integral decouples, as $\tilde
S_{ab}(k)$ is proportional to $1/(k^0)^2$ and the ratio of
$\tilde{S}$ functions is independent of the scale of $k^0$:
\begin{equation}
\begin{split}
 & \Delta U(\epsilon)  = \int\limits_{\epsilon \sqrt{s}/2}^{\sqrt{s}}
        \frac{d^3k}{{k}^0} \tilde{S}_{ab}(k)\big(1-w^R(k)\big)
     = -\int\limits_{\epsilon \sqrt{s}/2}^{\sqrt{s}} \frac{dk^0}{k^0}
       \int d\Omega_k F(a,b,c,d;\Omega_k)   
 =  (\gamma_R-\gamma)\log\frac{\epsilon}{2}
\\
& \gamma_R-\gamma =\int d\Omega_k F(a,b,c,d;\Omega_k)
\\
&F(a,b,c,d;\Omega_k) =      (k^0)^2
                      \tilde{S}_{ab}(k)
              \frac{
                \tilde{S}_{aC}(k) +\tilde{S}_{bD}(k) 
               +\tilde{S}_{aD}(k) +\tilde{S}_{bC}(k) }
              { \tilde{S}_{ab}(k) +\tilde{S}_{CD}(k) }.
\label{waga-vs2}
\end{split}
\end{equation}
The evaluation of $(\gamma_R-\gamma)$ simplifies
if we note that this integral is Lorentz-invariant. It is
easiest to show for the initial form of $\Delta U(\epsilon)$ given in Eq.\
(\ref{waga-vs}). The only apparently Lorentz-variant part of Eq.\
(\ref{waga-vs}) are the integration limits of $k^0$. Upon a
Lorentz transformation with an arbitrary parameter $\beta$, the $k_0$ 
transforms as 
\begin{eqnarray}
k_0\to {k_0}' = \frac{k_0 -\vec{\beta}\vec{ k}}{\sqrt{1-\vec{ \beta}^2}} 
     = k_0 A(\Omega_k);\;\;\;\;
A(\Omega_k) =\frac{1-|\vec{ \beta}|\cos\angle(\vec{\beta}\vec{k})}
                  {\sqrt{1-\vec{\beta}^2}} 
\end{eqnarray}
and the integral $\Delta U(\epsilon)$ becomes
\begin{equation}
 \Delta U(\epsilon) = \int\limits_{A^{-1}\epsilon \sqrt{s}/2}^{A^{-1}\sqrt{s}} 
\frac{d^3k'}{({k'}^0)^3} F(a',b',c',d';\Omega_{k'}).
\end{equation}
By the rescaling transformation $k' \to A k'$,
due to the identity 
$\Omega_{k'}=\Omega_{Ak'}$ and the scale invariance of the function
$F(a,b,c,d;\Omega_k)$, the 
Lorentz invariance of $\Delta U(\epsilon)$ becomes transparent.

After substituting the definition of $\tilde{S}$ from Eq.\
(\ref{stilde0}) into  
Eq.~(\ref{waga-vs2}), we get
\begin{eqnarray}
\int d\Omega_k F(\Omega_k) 
        &=&  I_1 +I_2  
\\
I_1 &=&  \int d\Omega_k   (k^0)^2
                      \tilde{S}_{ab}(k)
              \frac{ 
                \tilde{S}_{aC}(k)  
               +\tilde{S}_{aD}(k)  }
              { \tilde{S}_{ab}(k) +\tilde{S}_{CD}(k) } 
\\
    &=&    \int d\cos\theta d\phi  \frac{\alpha}{4\pi^2} 
             \frac{2(ab)}{(n_ka)} 
              \frac{ (ac)(n_kd) -(ad)(n_kc) }
                   { (ab)(n_kc)(n_kd) +(cd)(n_ka)(n_kb) }
\\
I_2 &=&  \int d\Omega_k   (k^0)^2
                      \tilde{S}_{ab}(k)
              \frac{ 
                                  \tilde{S}_{bD}(k) 
                                 +\tilde{S}_{bC}(k) }
              { \tilde{S}_{ab}(k) +\tilde{S}_{CD}(k) }
                 = I_1(a\leftrightarrow b, c\leftrightarrow d),
\label{i12}
\end{eqnarray}
where $n_k^\mu =k^\mu/k^0$.
The integral contains scalar products of $k$ with four different
four-momenta 
$a,\dots,d$. With the parametrization of 
$(n_k u) = u^0 -u^z \cos\theta -u^x \sin\theta\cos\phi 
                                      -u^y \sin\theta\sin\phi$ 
the $d\phi$ integral in $I_1$ is in principle solvable. However, in the LAB
frame, i.e.~in the CMS frame of $(a,b)$, the $(n_kc)(n_kd)$ product in
the denominator would lead to a fourth-order polynomial and its zeros
would have to be found numerically. This problem can be
avoided by changing the Lorentz frame to the CMS$(a,c)$, i.e.\ the Breit
frame of $(a,c)$! The $I_2$ integral would similarly be evaluated in
CMS$(b,d)$ and then, due to the Lorentz invariance of the integrals $I_i$
we can simply add $I_1(a_1,\dots,d_1)+I_2(a_2,\dots,d_2)$. 
One must only remember that in both integrals the $a,\dots,d$ vectors
became transformed to different values (different frames), 
which we indicated above as $a_1$, $a_2$, etc. 

In the CMS$(a,c)$ we have $(n_k a) = a^0 -a \cos\theta$ and
$(n_k c) = c^0 +c \cos\theta$, so that
\begin{eqnarray}
I_1 &=&    \int d\cos\theta \frac{\alpha}{2\pi^2} \frac{(ab)}{(n_ka)} 
          \int d\phi 
              \frac{ L \cdot n_k }
                   { M(\theta) \cdot n_k }
\\
   &=&     \int d\cos\theta \frac{\alpha}{2\pi^2} \frac{(ab)}{(n_ka)} 
          \int d\phi 
  \frac{L^0-L^z\cos\theta -L_x\sin\theta\cos\phi -L_y\sin\theta\sin\phi}
       {M^0-M^z\cos\theta -M_x\sin\theta\cos\phi -M_y\sin\theta\sin\phi}
          \;\;\;\;\;\;
\\
 L^\mu &=& (ac)d^\mu - (ad)c^\mu
\\
 M^\mu(\theta) &=& (ab)(n_kc)d^\mu +(cd)(n_ka)b^\mu.
\end{eqnarray}
The integral over $d\phi$ can now be solved with the help of a textbook
formula 
(cf.~e.g.~\cite{Gradshteyn:1994}  2.558)
\begin{eqnarray}
\label{rhyzik}
&&\int\limits_{-\pi}^{\pi} d\phi\frac{l_0+l_1\cos\phi+l_2\sin\phi}
                            {m_0+m_1\cos\phi+m_2\sin\phi}
\nonumber
\\
   &=&2\pi\frac{l_1m_1 +l_2m_2}{m_1^2+m_2^2} 
 +\frac{2\pi}{\sqrt{m_0^2-m_1^2-m_2^2}} 
       \left(l_0 +\frac{l_1m_1 +l_2m_2}{m_1^2+m_2^2} m_0\right);
\;\;\;m_0^2 > m_1^2+m_2^2 \;\;\;\;\;
\\
  &=&2\pi l_0/m_0; \;\;\; m_1=m_2=0,
\end{eqnarray}
to give
\begin{equation}
I_1 =    \int d\cos\theta \frac{\alpha}{\pi} \frac{(ab)}{(n_ka)} 
\left(
   \frac{L_xM_x +L_yM_y}{M_x^2+M_y^2} 
 +\frac{L^0-L^z\cos\theta 
                -\frac{L_xM_x +L_yM_y}{M_x^2+M_y^2}(M_0-M_z\cos\theta)}
           {\sqrt{(M_0-M_z\cos\theta)^2 -(M_x^2+M_y^2)\sin^2\theta}} 
\right).
\label{finali1}
\end{equation}
There is one technical point related to Eq.~(\ref{finali1}), concerning the
positiveness of the quantity under the square root in the last term. 
It is guaranteed, however, by the 
positiveness of the denominator $M(\theta) \cdot n_k$, which is a sum of
dot products of time- and light-like four-vectors: at its minimum (with
respect to $\phi$) the $M(\theta) \cdot n_k$ is equal to
the quantity in question. 
The remaining $d\cos\theta$ integral we perform numerically.

The integral $I_2$ is evaluated analogously to $I_1$.

\newpage

\newpage
%
%

\newpage
%
%

\newpage
%
%

\newpage
%
%

\newpage
%
%
\newpage
%
%

\newpage
%
%

%
%

\end{document}